\documentclass[english,12pt]{article}
\usepackage{array}
\usepackage{graphicx}
\usepackage{amssymb}
\usepackage[english]{babel}
\usepackage{amsmath}
\usepackage{multirow}
\usepackage{prettyref}
\usepackage{babel}
\usepackage{units}
\usepackage[latin1]{inputenc}
\usepackage{amsfonts}
\usepackage{amssymb}
\usepackage{babel}
\usepackage{color}
\usepackage{cite}
\def\@fmsl@sh#1#2#3{\m@th\ooalign{$\hfil#1\mkern#2/\hfil$\crcr$#1#3$}}
 \def\eq#1\en{\begin{equation}#1\end{equation}}
\def\s[#1,#2]{[#1\stackrel{\star}{,}#2]}
\def\sx[#1,#2]{[#1\stackrel{\star_{x}}{,}#2]}

\newcommand{\nc}{\newcommand}
\nc{\beq}{\begin{equation}}
\nc{\eeq}{\end{equation}}
\nc{\beqa}{\begin{eqnarray}}
\nc{\eeqa}{\end{eqnarray}}

\def\bc{\begin{center}}
\def\ec{\end{center}}

\def\gsim{\mathrel{\mathpalette\atversim>}}

\def\bc{\begin{center}}
\def\ec{\end{center}}

\def\gsim{\mathrel{\rlap{\lower4pt\hbox{\hskip1pt$\sim$}}

    \raise1pt\hbox{$>$}}}       

\def\gsim{\mathrel{\rlap{\lower4pt\hbox{\hskip1pt$\sim$}}
    \raise1pt\hbox{$>$}}}       

\begin{document}
\makeatletter
\def\fmslash{\@ifnextchar[{\fmsl@sh}{\fmsl@sh[0mu]}}
\def\fmsl@sh[#1]#2{%
  \mathchoice
    {\@fmsl@sh\displaystyle{#1}{#2}}%
    {\@fmsl@sh\textstyle{#1}{#2}}%
    {\@fmsl@sh\scriptstyle{#1}{#2}}%
    {\@fmsl@sh\scriptscriptstyle{#1}{#2}}}
\def\@fmsl@sh#1#2#3{\m@th\ooalign{$\hfil#1\mkern#2/\hfil$\crcr$#1#3$}}
\makeatother

\thispagestyle{empty}
\begin{titlepage}
\boldmath
\begin{center}
  \Large {\bf Vanishing of Quantum Gravitational Corrections \\ to Vacuum Solutions of General Relativity \\ at Second Order in Curvature}
    \end{center}
\unboldmath
\vspace{0.2cm}
\begin{center}
{  {\large Xavier Calmet}\footnote{x.calmet@sussex.ac.uk}$^{, a, b}$}
 \end{center}
\begin{center}
$^a${\sl Department of Physics and Astronomy, 
University of Sussex, Brighton, BN1 9QH, United Kingdom
}\\
$^b${\sl PRISMA Cluster of Excellence and Mainz Institute for Theoretical Physics, Johannes Gutenberg University, 55099 Mainz, Germany }\\
\end{center}
\vspace{5cm}
\begin{abstract}
\noindent
In this note we prove that quantum gravitational corrections to vacuum solutions of Einstein's equations vanish at second order in curvature.
\end{abstract}  
\vspace{5cm}
\end{titlepage}



\newpage

Exact solutions, and in particular vacuum solutions, play an important role in general relativity, see e.g. \cite{Stephani:2003tm}. Using effective theory techniques, we prove in this note that quantum gravitational corrections to vacuum solutions of Einstein's equations vanish at second order in curvature.

Deriving an effective action for quantum gravity requires starting from general relativity and integrating out fluctuations of the graviton and of any matter fields that are irrelevant for the problem under consideration. Doing so, see e.g. \cite{Donoghue:2017pgk} for a recent review, one obtains a classical effective action given at second order in curvature by:
\begin{eqnarray}\label{action1}
S_{local} &=& \int d^4x \, \sqrt{-g} \left[  \frac{1}{2}  M_P^2   \mathcal{R} + c_1 \mathcal{R}^2 + c_2 \mathcal{R}_{\mu\nu}\mathcal{R}^{\mu\nu}+ c_4   \Box \mathcal{R}   \right],
\end{eqnarray}
for the local part and for the non-local part by
\begin{eqnarray}
S_{NL} &=& \int d^4x \, \sqrt{-g} \left[ - b_1 \mathcal{R} \log \frac{\Box}{\mu^2_1}\mathcal{R} - b_2 \mathcal{R}_{\mu\nu}  \log \frac{\Box}{\mu^2_2}\mathcal{R}^{\mu\nu}  
- b_3 \mathcal{R}_{\mu\nu\rho\sigma}  \log \frac{\Box}{\mu^2_3}\mathcal{R}^{\mu\nu\rho\sigma}  \right],
\end{eqnarray}
where $\mathcal{R}$, $\mathcal{R}_{\mu\nu}$ and $\mathcal{R}_{\mu\nu\rho\sigma}$ are respectively the Ricci scalar, the Ricci tensor and the Riemann tensor.  The scales $\mu_i$ are renormalization scales. The term $\Box \mathcal{R}$ is a total derivative and thus does not contribute to the equation of motions. The effective action is a derivative expansion since the Ricci scalar, Ricci tensor and Riemann tensor contain two derivatives of the metric. However, as the invariant terms are constructed using the Ricci scalar, the Ricci tensor and the Riemann tensor,  it also corresponds to a curvature expansion. In this approach the terms curvature expansion and derivative expansion are thus used interchangeably.  The expansion of the effective action here is analogous to that done in the case of chiral perturbation theory. The non-local action is obtained by integrating out massless fluctuations. One could also integrate out massive fields, which would generate operators of the type $\mathcal{R}m^4 \Box^{-1} \mathcal{R}$ (see e.g. \cite{Codello:2015mba}), where $m$ is the mass, at the second order in the derivative expansion. We are restricting our considerations to massless fields such as the graviton, but the generalization to massive fields is straightforward.

This effective action is obtained by quantizing general relativity directly. It is well known that this quantum field theory is not renormalizable, but as argued by Donoghue years ago  \cite{Donoghue:1993eb} this is not an issue as it simply means that new measurements are needed order by order in the curvature/derivative expansion of the effective action to determine the Wilson coefficients that are not calculable due to the non-renormalizability of quantized general relativity. However, this does not mean that predictions are not possible as the Wilson coefficients of the non-local operators such as the $\log \Box$ terms considered here are calculable. Their values are given by the number of massless fields that are integrated out. In that sense, any ultra-violet complete theory of quantum gravity that has general relativity in its low energy regime (which obviously applies to string theory), will be described by this effective action. Obviously fields specific to a given ultra-violet completion such as e.g. the Kaluza-Klein modes of the graviton would have to be included. The effective theory approach is thus a useful framework to consider quantum gravitational corrections to Einstein's theory of general relativity when considering physics at energies below the Planck scale. 

Our aim is to study quantum gravitational corrections to Einstein's field equations. In particular, we are interested in vacuum solutions such as black hole solutions. It is well known that the variation of the local part of the action does not lead to any correction at second order in curvature as the variations with respect to the metric of the two second order terms in curvature are proportional to either $R$ or $R_{\mu\nu}$ which both vanish in vacuum. We can thus focus on the non-local part of the action. Varying this part of the action we find:
\begin{eqnarray}
\frac{\delta S_{NL}}{\delta g^{\alpha\beta}} = \frac{\delta}{\delta g^{\alpha\beta}}\Bigg[ \int d^4x \, \sqrt{-g} \left[ - b_1 \mathcal{R} \log \frac{\Box}{\mu^2_1}\mathcal{R} - b_2 \mathcal{R}_{\mu\nu}  \log \frac{\Box}{\mu^2_2}\mathcal{R}^{\mu\nu}  
- b_3 \mathcal{R}_{\mu\nu\rho\sigma}  \log \frac{\Box}{\mu^2_3}\mathcal{R}^{\mu\nu\rho\sigma}  \right] \Bigg]. \nonumber \\ 
\end{eqnarray}
At this stage we need to specify how to apply the $\log(\Box/\mu^2)$ on the different terms in the non-local action. This differential operator can be defined in terms of an integral representation:
\begin{eqnarray}
\log \frac{\Box}{\mu^2}= \int_0^\infty ds \Bigg ( \frac{1}{\mu^2+s} - \frac{1}{\Box +s} \Bigg).
\end{eqnarray}
It has been shown \cite{Donoghue:2014yha,Codello:2015pga} that the variation $\delta/\delta g^{\alpha\beta} \log(\Box/\mu^2)$ leads to terms in the field equations that are higher order than two in curvature. We can thus safely ignore this variation as we are considering corrections to the field equations to second order in curvature. Using the integral representation of the $\log$, we can now calculate:
\begin{eqnarray}
\log \frac{\Box}{\mu^2} \Big(\mathcal{R}_{\mu\nu\rho\sigma}\mathcal{R}^{\mu\nu\rho\sigma}\Big) &=& \frac{1}{2} \log \frac{\Box}{\mu^2}\Big( \mathcal{R}_{\mu\nu\rho\sigma}\Big)\mathcal{R}^{\mu\nu\rho\sigma}+\frac{1}{2} \mathcal{R}_{\mu\nu\rho\sigma}\log \frac{\Box}{\mu^2}\Big( \mathcal{R}^{\mu\nu\rho\sigma}\Big),
\end{eqnarray}
which is valid to second order in curvature. We used a Taylor series expansion for the fraction containing the $\Box$-operator. It is then easy to see that the deviation from the distributivity property involves derivative terms which imply terms of higher curvature than two. Another way to see this is to realize that the $\log \Box/\mu^2$ operator stands for a non-local function $L(x,y,\mu)$ (see e.g. \cite{Calmet:2017qqa}) which could be calculated, for example, using a Riemann normal coordinates expansion \cite{BD}, in which propagators are expanded in powers of the curvature/derivative. However, since the action at second order in curvature is by definition quadratic in the curvature, we can drop any derivative term appearing in $L(x,y,\mu)$ at this order in curvature and safely commute the Riemann tensors with the $\log \Box/\mu^2$ operator at second order in curvature.

 As long as we restrict our considerations to second order corrections in curvature we can simplify the right-hand side further (passing metrics to lower and raise indices past the $\log$ operator will simply generate higher order derivative terms), we get
\begin{eqnarray}
\log \frac{\Box}{\mu^2} \Big(\mathcal{R}_{\mu\nu\rho\sigma}\mathcal{R}^{\mu\nu\rho\sigma}\Big) &=& \mathcal{R}_{\mu\nu\rho\sigma}\log \frac{\Box}{\mu^2}\Big( \mathcal{R}^{\mu\nu\rho\sigma}\Big).
\end{eqnarray}
We can now express $ \mathcal{R}_{\mu\nu\rho\sigma}\mathcal{R}^{\mu\nu\rho\sigma}$ in terms of the standard Gauss-Bonnet identity and find:
\begin{eqnarray}
\mathcal{R}_{\mu\nu\rho\sigma}\log \frac{\Box}{\mu^2}\Big( \mathcal{R}^{\mu\nu\rho\sigma}\Big)=\log \frac{\Box}{\mu^2} \Big(4 \mathcal{R}_{\mu\nu}\mathcal{R}^{\mu\nu}- \mathcal{R}^2 \Big)=4 \mathcal{R}_{\mu\nu} \log \frac{\Box}{\mu^2}  \mathcal{R}^{\mu\nu}- \mathcal{R}\log \frac{\Box}{\mu^2} \mathcal{R}.
\end{eqnarray}
which is the non-local version of the Gauss-Bonnet identity which is however only valid to second order in curvature.
The variation of the non-local part action thus becomes:
\begin{eqnarray}
\frac{\delta}{\delta g^{\alpha\beta}}S_{NL} &=&  \int d^4x \,  \frac{\delta}{\delta g^{\alpha\beta}}\big(\sqrt{-g}\big) \left[ - b_1\mathcal{R} \log \frac{\Box}{\mu^2_1}\mathcal{R} - b_2 \mathcal{R}_{\mu\nu}  \log \frac{\Box}{\mu^2_2}\mathcal{R}^{\mu\nu}  
- b_3 \mathcal{R}_{\mu\nu\rho\sigma}  \log \frac{\Box}{\mu^2_3}\mathcal{R}^{\mu\nu\rho\sigma}  \right]  \nonumber \\ 
&&+  \int d^4x \,  \sqrt{-g} \Big[ - 2 b_1\frac{\delta}{\delta g^{\alpha\beta}}\big(\mathcal{R}\big) \log \frac{\Box}{\mu^2_1}\mathcal{R} - 2 b_2 \frac{\delta}{\delta g^{\alpha\beta}}\big(\mathcal{R}_{\mu}^{\nu}\big)  \log \frac{\Box}{\mu^2_2}\mathcal{R}^{\mu}_{\nu}  
\nonumber \\ 
&&
- b_3  \log \frac{\Box}{\mu^2_3} \Big ( 4 \frac{\delta}{\delta g^{\alpha\beta}}\big(\mathcal{R}_{\mu}^{\nu}\big)\mathcal{R}^{\mu}_{\nu}- \frac{\delta}{\delta g^{\alpha\beta}}\big(\mathcal{R}\big) \mathcal{R}  \Big) \Big]. 
\end{eqnarray}
It is straightforward to see that all these terms are either proportional to $\mathcal{R}$ or $\mathcal{R}_{\mu\nu}$ which both vanish in vacuum. 

We have thus shown that there are no second order quantum gravitational corrections to vacuum solutions. This applies to all vacuum solutions in general relativity, e.g. black hole solutions. This proof confirms the results obtained for the Schwarzschild solution in \cite{Calmet:2017qqa}. Note that while this previous result was obtained by doing a lengthy calculation, the proof presented here is rather short and completely generic. 

A corollary to our result is that the effective action for quantum gravity at second order in curvature can be written as
\begin{eqnarray}
S&=& \int d^4x \, \sqrt{-g} \left(  \frac{1}{2}  M_P^2   \mathcal{R} + c_1 \mathcal{R}^2 + c_2 \mathcal{R}_{\mu\nu}\mathcal{R}^{\mu\nu}
 - \bar b_1 \mathcal{R} \log \frac{\Box}{\mu^2_1}\mathcal{R} - \bar b_2 \mathcal{R}_{\mu\nu}  \log \frac{\Box}{\mu^2_2}\mathcal{R}^{\mu\nu}     \right),
\end{eqnarray}
with $\bar b_1= b_1-b_3$ and  $\bar b_2=b_2+4 b_3 $ when ignoring topological terms. It implies that in quantum general relativity, there are just four independent Wilson coefficients at second order in curvature. We remind the reader that while the $c_i$ cannot be predicted from first principles, the $b_i$ can be calculated in a model independent manner, see e.g. \cite{Donoghue:2014yha}. In this compact form, it is easy to identify directly from the action the masses of the massive spin-2 and spin-0 fields present in the action \cite{Calmet:2014gya}.

Let us finally emphasize that our result by no means implies that there are no new intrinsic solutions to higher-derivative gravity. Such solutions do exist as shown by Lu et al. in \cite{Lu:2015cqa}, we simply find that that are no quantum gravitational corrections to vacuum solutions of general relativity at second order in curvature.

{\it Acknowledgments:}
The work of XC is supported in part  by the Science and Technology Facilities Council (grant number ST/P000819/1). XC is very grateful to MITP for their generous hospitality during the academic year 2017/2018. 


\bigskip{}


\baselineskip=1.6pt 

\begin{thebibliography}{10}

\bibitem{Stephani:2003tm} 
  H.~Stephani, D.~Kramer, M.~A.~H.~MacCallum, C.~Hoenselaers and E.~Herlt,
  ``Exact solutions of Einstein's field equations,'' Cambridge University Press, 2003, 701 pp,
  doi:10.1017/CBO9780511535185
  
\bibitem{Donoghue:2017pgk} 
  J.~F.~Donoghue, M.~M.~Ivanov and A.~Shkerin,
  arXiv:1702.00319 [hep-th].
  
\bibitem{Codello:2015mba} 
  A.~Codello and R.~K.~Jain,
  Class.\ Quant.\ Grav.\  {\bf 33}, no. 22, 225006 (2016)
  doi:10.1088/0264-9381/33/22/225006
  [arXiv:1507.06308 [gr-qc]].
  
\bibitem{Donoghue:1993eb} 
  J.~F.~Donoghue,
  Phys.\ Rev.\ Lett.\  {\bf 72}, 2996 (1994)
  doi:10.1103/PhysRevLett.72.2996
  [gr-qc/9310024].
  
\bibitem{Donoghue:2014yha} 
  J.~F.~Donoghue and B.~K.~El-Menoufi,
  Phys.\ Rev.\ D {\bf 89}, no. 10, 104062 (2014)
  doi:10.1103/PhysRevD.89.104062
  [arXiv:1402.3252 [gr-qc]].
  
    
\bibitem{Codello:2015pga} 
  A.~Codello and R.~K.~Jain,
  Class.\ Quant.\ Grav.\  {\bf 34}, no. 3, 035015 (2017)
  doi:10.1088/1361-6382/aa549d
  [arXiv:1507.07829 [astro-ph.CO]].
  
\bibitem{Calmet:2017qqa} 
  X.~Calmet and B.~K.~El-Menoufi,
  Eur.\ Phys.\ J.\ C {\bf 77}, no. 4, 243 (2017)
  doi:10.1140/epjc/s10052-017-4802-0
  [arXiv:1704.00261 [hep-th]].
  
\bibitem{BD} N.~D.~Birrell and P.~C.~W.~Davies, ``Quantum Fields in Curved Space,'' (Cambridge University Press, Cambridge, 1982).

\bibitem{Calmet:2014gya} 
  X.~Calmet,
  Mod.\ Phys.\ Lett.\ A {\bf 29}, no. 38, 1450204 (2014)
  doi:10.1142/S0217732314502046
  [arXiv:1410.2807 [hep-th]].
  
\bibitem{Lu:2015cqa} 
  H.~Lu, A.~Perkins, C.~N.~Pope and K.~S.~Stelle,
  Phys.\ Rev.\ Lett.\  {\bf 114}, no. 17, 171601 (2015)
  doi:10.1103/PhysRevLett.114.171601
  [arXiv:1502.01028 [hep-th]].
 
\end{thebibliography}
\end{document}